\documentclass[A4paper]{article}
\usepackage[dvips]{graphicx}
\usepackage{enumerate}
\begin{document}
%&amslplain

\pagestyle{plain} 
\setcounter{page}{1}
\setlength{\topmargin}{-40pt}
\setlength{\headheight}{0pt}
\setlength{\marginparwidth}{-10pt}

\title{A Proposal for Quantum Key Distribution using Quantum Games}
\author{Norihito Toyota, %\and 
Hokkaido Information University, Ebetsu, Nisinopporo 59-2, Japan \and email :toyota@do-johodai.ac.jp }
\date{}
\maketitle
\begin{abstract}
We  study a new QKD that is different from the scheme proposed by \cite{Ramz2}, though it essentially takes our ground on three-player  quantum games and Greenberg-Horne-Zeilinger triplet entangled state (GHZ state) \cite{Gree} is used. 
In the scheme proposed in this paper,  players in the game, Bob and Charlie (and Alice also) can  get some common key or information (applied strategies and their payoffs in the game),  when Alice informs Bob and Charlie about some results of the measurement made by her. 
Even if somebody else knows the public information,  he/she can not get any key information.  
There is not any arbiter in our scheme, since existence of an arbiter increases the risk of wiretapping. 
%For  it is difficult to detect wiretapping,  when an arbiter repeatedly sends classical information.    
Lastly we discuss robustness of the proposed QKD method for eavesdrop. 
We show that though maximally entangled case and  non-entangled case essentially provide an equivalent way as QKD,  the latter is not available in the case where there are some eavesdroppers. 
At the same time, we point put that the entanglement of the initial state is crucial when a partially entangled state is used.   
 \end{abstract}
\begin{flushleft}
\textbf{keywords:}
Three-Player Quantum Game, Quantum Key Distribution, Payoff, Entangled State,  Eavesdrop
\end{flushleft}

\section{Introduction}
\hspace{5mm} 
Quantum version of information science has opened the doors of  new and large possibility of computer science\cite{Niel}.  
 Quantum game\cite{Meyer} is an interesting subject and remains in the realm of the unknown in potential capacities. 
 Two schemes for quantum games mainly have been  proposed so far.  
 One has been proposed by Eisert et al. \cite{Eise} where the strategy space of players is a two parameter set of $2 \times 2 $ matrices and Prisoners Dilemma was discussed. 
It is shown that starting with  maximally entangled initial state, the dilemma disappears for a suitable quantum strategy. 
Moreover they pointed out that a quantum strategy displays its superiority to all the  classical strategies. 
The details of this scheme were also reviewed by Rosero \cite{Rose}.   
 Another one has been  introduced by Marinatto and Weber \cite{Mari} applying it to the game of Battle of Sexes.  
 In their scheme, starting with  maximally entangled initial state,  the players are allowed to apply the probabilistic tactics of unitary operators.  
 As result, they found the strategy for which  both the players can get equal payoffs. 
Recently two schemes could be studied by a unified way\cite{Nawa1}. 
Two schemes are one aspect of the generalized quantization scheme developed by Nazawa and Toor\cite{Nawa1}.  
Furthermore, the scheme was  extended to three-player quantum games  by Ramzan and Khan\cite{Ramz1}. 

  Quantum game theory is not only a tool to  resolve a dilemma in some games and  find better payoffs than ones of classical strategies, but also it has more potential ability and will provide wide types of communication protocols. 
For further details, you can consult the article given by Iqbal\cite{Iqbal}.  
In fact  Ramzan and Khan studied an interesting aspects of communication based on  three-player  quantum Prisoner's Dilemma\cite{Ramz1}. 
Moreover the authors  investigated a cryptographic protocol based on a scheme of the generalized three-player quantum game\cite{Ramz2}. 

In classical cryptography in 20-th century, key distribution that generates a private key in a secure way between two or several remote parties  is an important subject. 
In public key crypto-systems such as Rivest-Shamir-Adelman(RSA)\cite{Rive}, the receiver generates a pair of keys; a public key and a private key. 
The security of the communication relies on the difficulty to factorize a large integer into some prime numbers. 
The public key is used to encrypt the message by a sender, while the private one is used for a receiver to decrypt   it.        
But it has been proved that quantum algorithm can solve the factorization problem in polynomial time\cite{Shor}. 
On the other hand, quantum information theory itself provides some ways of quantum key distribution (QKD). 
First protocol was proposed by Bennet and Brassard\cite{Benn}. 
After that Eckert proposed a different approach to QKD\cite{Ecke}.  
Many protocols are developed as of today\cite{Vale}.    

Inspired by a series of researches, we propose a new kind of communication scheme based on  three-player  quantum game in this paper.  
We mainly study QKD that is different from the scheme proposed \cite{Ramz2}, though it essentially takes our ground on three-player  quantum games and Greenberg-Horne-Zeilinger triplet entangled state (GHZ state) \cite{Gree} is used. 
In the scheme of  \cite{Ramz2}, Alice finds messages sent from Bob and Charlie by making a measurement of the qubits that are manipulated by their unitary operators. 
In our scheme proposed in this paper, Bob and Charlie (and Alice also) get some common key or information 
by knowing the some information of the measurement made by Alice.  
In \cite{Ramz1} where the quantum version of Prisoner's Dilemma game  is adopted, Bob and Charlie can extract  information about the strategy applied by Alice from their payoffs  by mutual understanding that they will apply the same strategy.  
This protocol stands as an information communication rather than  QKD because this assumes the mutual  understanding. 
The information about payoffs in \cite{Ramz1}  is brought by an arbiter.  
On the contrary there are not any arbiters in our scheme. 
Existence of an arbiter increase the risk of wiretapping, because when an arbiter sends classical information, but not quantum state, to Bob and Charlie,  it is difficult to detect wiretapping. 
As it is better that there are not any arbiters in a protocol in view  of wiretapping,   
an arbiter is excluded in our scheme.
In the proposed protocol here, a scheme that Bob and Charlie (and Alice) can get some information (key) by results obtained from Alice's experiment and it can be made in the secrecy from other people.
Thus it provides a sort of QKD.   
In fact Bob and Charlie (and Alice) can extract other people's full information (applied strategies and their payoffs)  theoretically by the information shown by Alice.    
We show that methods using an entangled state and a non-entangled state essentially equivalent as QKD but a method using partially entangled state give a
different one from the formers in the protocol. 

Lastly we discuss robustness for eavesdrop. 
We show that though maximally entangled case and  non-entangled case provide an essentially equivalent way as QKD,  
the latter is not available in the case there are any eavesdroppers.

\section{Framework of Quantum Game}
\hspace{5mm} 
The protocol proposed in this paper is based on  three-player quantum games. 
Three players are Alice , Bob and Charlie following custom in this field.  
As result of the game, three player find other players' strategies and payoffs after Alice revealed some information to Bob and  Charlie about the result of the quantum game. 
First of all we describe the framework of quantum games.

Basically we follow the generalized formalism of quantum games proposed by Nawaz and Toor \cite{Nawa1} and its extension given by Ramzan and Khan \cite{Ramz1}. 
In  scheme of this paper, Alice, Bob and Charlie can choose one  among two strategies F and T, respectively.  
We  express the strategies F as 0 and T as 1, respectively and a set of strategies of three players as (0,0,1) for example. 
The set of strategies (0,0,1) denotes that Alice  applies the strategy F, Bob does F and Charlie does T.  

First Alice prepares an initial quantum state that consists of three qubits and passes second qubit and third one  to Bob and Charlie, respectively.  
Bob and Charlie accept one qubit, respectively  and Alice keeps remaining one qubit that is first qubit.   
We suppose the initial quantum state shared among three players is the generalized GHZ state;
\begin{equation}
|\psi_{in}>= \cos\frac{\gamma}{2}|000> + i\sin \frac{\gamma}{2}|111>,
\end{equation}
where $0\geq \gamma \geq \pi/2$.  
There is no entanglement for $\gamma=0$ and the case of $\gamma=\pi/2$ denotes maximally entangled state that has the largest von Neumann entropy. 

Next the players locally manipulate their individual qubits by some unitary operators; Alice, Bob and Charlie make the unitary transformations on the first qubit, 
the second qubit and third qubit, respectively.    
The classical strategies F and T are assigned to the two basis vectors $|0>$ and $|1>$ in the Hilbert space, respectively. 
The strategies of the players are represented by the unitary operator $U_k$ defined by \cite{Nawa1} 
\begin{equation}
U_k=\cos\frac{\theta_k}{2} R_k + \sin \frac{\theta_k}{2}Q_k, 
\end{equation}
where $k=$ A(Alice), B(Bob) and C(Charlie), and $R_k$ and $Q_k$ are unitary operators defined by 
\begin{eqnarray}
&R_k|0>=e^{i\alpha_k}|0>, &R_k|1>=e^{-i\alpha_k}|1>, \nonumber \\
&Q_k|0>=e^{i(\frac{\pi}{2}-\beta_k)}|1>, &Q_k|1>=e^{i(\frac{\pi}{2}+\beta_k)}|0>, 
\end{eqnarray}
where  $0 \leq \theta_k \leq \pi,$ and $ -\pi\leq \alpha_k, \beta_k \leq \pi$. 
After applying the local operators of three players, the density matrix of the initial state $\rho_{in}= |\psi_{in}><\psi_{in}|$  changes to  
\begin{equation} 
\rho_f = ( U_A\bigotimes U_B\bigotimes U_C ) \rho_{in} ( U_A\bigotimes U_B\bigotimes U_C )^\dagger. 
\end{equation}

After that, Bob and Charlie return their qubits to Alice and so Alice gets $\rho_{f}$. 
To determine the payoffs for three players, we introduce the following payoff operator as a measurement operator; 
\begin{equation}
\$^{(k)}  =\$^{(k)}_{000}P_{000}+ \$^{(k)}_{001}P_{001}+\$^{(k)}_{010}P_{010}+\$^{(k)}_{100}P_{100}+\$^{(k)}_{011}P_{011}+\$^{(k)}_{110}P_{110}+\$^{(k)}_{101}P_{101}+\$^{(k)}_{111}P_{111},
\end{equation}
where
\begin{eqnarray}
P_{000}&=& |\psi_{000}>< \psi_{000}|, \;\;\;\;\;  |\psi_{000}>= \cos\frac{\delta}{2}|000>+i\sin \frac{\delta}{2}|111>,\nonumber  \\
P_{111}&=& |\psi_{111}>< \psi_{111}|, \;\;\;\;\;  |\psi_{111}>= \cos\frac{\delta}{2}|111>+i\sin \frac{\delta}{2}|000>,\nonumber  \\
P_{001}&=& |\psi_{001}>< \psi_{001}|, \;\;\;\;\;  |\psi_{001}>= \cos\frac{\delta}{2}|001>+i\sin \frac{\delta}{2}|110>,\nonumber  \\
P_{110}&=& |\psi_{110}>< \psi_{110}|, \;\;\;\;\;  |\psi_{110}>= \cos\frac{\delta}{2}|110>+i\sin \frac{\delta}{2}|001>,\nonumber  \\
P_{010}&=& |\psi_{010}>< \psi_{010}|, \;\;\;\;\;  |\psi_{010}>= \cos\frac{\delta}{2}|010>-i\sin \frac{\delta}{2}|101>,\nonumber  \\
P_{101}&=& |\psi_{101}>< \psi_{101}|, \;\;\;\;\;  |\psi_{101}>= \cos\frac{\delta}{2}|101>-i\sin \frac{\delta}{2}|010>,\nonumber  \\
P_{011}&=& |\psi_{011}>< \psi_{011}|, \;\;\;\;\;  |\psi_{011}>= \cos\frac{\delta}{2}|011>-i\sin \frac{\delta}{2}|100>,\nonumber  \\
P_{100}&=& |\psi_{100}>< \psi_{100}|, \;\;\;\;\;  |\psi_{100}>= \cos\frac{\delta}{2}|100>-i\sin \frac{\delta}{2}|011>
\end{eqnarray}
with $ 0\leq \delta \leq \pi/2$ and $ \$_{abc}^{(k)} $ are the elements of the payoff matrix given in table 1. 
$\delta$ denotes the degree of entanglement in the base for measurement (computational) base.  
In quantum version of usual game theories, a payoff is really given by an expectation value. 
In this scheme, Alice takes the final projective measurement (von Neumann measurement) in the computational basis given by Eq.(6).  
The expected payoffs for three players are obtained as the mean values of the payoff operators;
\begin{equation}
P^k  (\theta_k, \alpha_k,\beta_k, \delta,\gamma )= Tr ( \$^{(k)}_{abc} \rho_f ),
\end{equation}
where $a,b,c\in\{0,1\}$ that mean strategies of three players and $Tr$ is taken for above $ |\psi_{abc}>$ basis.  
 By using equations (1)$\sim$(7), the expected payoffs for three players are given by

\begin{eqnarray}
P^{k} (\theta_k, \alpha_k,\beta_k,\delta,\gamma )&=&  C_AC_BC_C \Big( \eta_1 \$^{(k)}_{000} + \eta_2 \$^{(k)}_{111} +\xi  (\$^{(k)}_{000}-\$^{(k)}_{111} ) \cos2(\alpha_A +\alpha_B+\alpha_C ) \Big) \nonumber \\
&+& S_AS_BS_C \Big( \eta_2 \$^{(k)}_{000} + \eta_1 \$^{(k)}_{111} -\xi  (\$^{(k)}_{000}-\$^{(k)}_{111} ) \cos2(\beta_A +\beta_B+\beta_C ) \Big) \nonumber\\
&+& C_AC_BS_C\Big( \eta_1 \$^{(k)}_{001} + \eta_2 \$^{(k)}_{110} +\xi  (\$^{(k)}_{001}-\$^{(k)}_{110} ) \cos2(\alpha_A +\alpha_B-\beta_C ) \Big) \nonumber\\
&+& S_AS_BC_C \Big( \eta_2 \$^{(k)}_{001} + \eta_1 \$^{(k)}_{110} -\xi  (\$^{(k)}_{001}-\$^{(k)}_{110} ) \cos2(\beta_A +\beta_B-\alpha_C ) \Big) \nonumber\\
&+& S_AC_BC_C \Big( \eta_1 \$^{(k)}_{100} + \eta_2 \$^{(k)}_{011} -\xi  (\$^{(k)}_{100}-\$^{(k)}_{011} ) \cos2(-\beta_A +\alpha_B+\alpha_C ) \Big) \nonumber\\
&+& C_AS_BS_C \Big( \eta_2 \$^{(k)}_{100} + \eta_1 \$^{(k)}_{011} +\xi  (\$^{(k)}_{100}-\$^{(k)}_{011} ) \cos2(-\alpha_A +\beta_B-\beta_C ) \Big) \nonumber\\
&+& S_AC_BS_C \Big( \eta_1 \$^{(k)}_{101} + \eta_2 \$^{(k)}_{010} -\xi  (\$^{(k)}_{101}-\$^{(k)}_{010} ) \cos2(\beta_A -\alpha_B+\beta_C ) \Big) \nonumber\\
&+& C_AS_BC_C\Big( \eta_2 \$^{(k)}_{101} + \eta_1 \$^{(k)}_{010} +\xi  (\$^{(k)}_{101}-\$^{(k)}_{010} ) \cos2(\alpha_A -\beta_B+\alpha_C ) \Big) \nonumber\\
&+&\frac{1}{8} \sin[\theta_A,\theta_B,\theta_C ] \times \nonumber\\
&\Bigl\{& \cos\delta \sin\gamma \cos (\alpha_A +\alpha_B+\alpha_C-\beta_A -\beta_B-\beta_C ) 
\sum_{a,b,c \in \{0,1 \} } \$_{abc}(-1)^{(a+b+c)} \nonumber\\
&-&\cos\gamma \sin\delta \Bigl( 
+\bigl( \$^{(k)}_{000}-\$^{(k)}_{111} \bigr) \cos(\alpha_A +\alpha_B+\alpha_C +\beta_A +\beta_B+\beta_C)   \nonumber\\
&&+\bigl( \$^{(k)}_{110}-\$^{(k)}_{001} \bigr) \cos(\alpha_A +\alpha_B-\alpha_C +\beta_A +\beta_B-\beta_C)   \nonumber\\
&&+\bigl( \$^{(k)}_{010}-\$^{(k)}_{101} \bigr) \cos(\alpha_A -\alpha_B+\alpha_C +\beta_A -\beta_B+\beta_C)   \nonumber\\
&&+\bigl( \$^{(k)}_{100}-\$^{(k)}_{011} \bigr) \cos(\alpha_A -\alpha_B-\alpha_C +\beta_A -\beta_B-\beta_C)   
\Bigr)   \Bigr\}, 
\end{eqnarray}
where 
\begin{eqnarray}
C_k&=&\cos^2 (\theta_k/2), \;\;\;S_k= \sin^2(\theta_k/2),\;\;\;\;and \,\, so\;\; C_k+S_k=1 \\
\eta_1&=&\cos^2\frac{\gamma}{2}\cos^2\frac{\delta}{2}+ \sin^2\frac{\gamma}{2}\sin^2\frac{\delta}{2},\\
\eta_2&=&\sin^2\frac{\gamma}{2}\cos^2\frac{\delta}{2}+ \cos^2\frac{\gamma}{2}\sin^2\frac{\delta}{2},\\
\xi &=& \frac{1}{2} \sin(\delta)\sin(\gamma),\\
\sin[\theta_A,\theta_B,\theta_C ] &= &\sin(\theta_A)\sin(\theta_B)\sin(\theta_C). 
\end{eqnarray}
This is a function of the entanglement parameters $\gamma$ and $\delta$, strategy parameters of three players, $\alpha_k$, $\beta_k$ and $\theta_k$,  
and the elements $ \$ ^{(k)}_{abc} $of a payoff matrix. 
%%%%%%%%%% $\$$
New parameters $\eta_1$, $\eta_2$ and $\xi $ are essentially entangled parameters, and  $\theta_k$  
 means the strategy parameter for player $k$.

\begin{center}%%%%%%%%%%%%%%%%%%%%%%%%%%%%%
Table 1. The payoff matrix for a three-player game where the first number in the parenthesis 
denotes the \hspace*{6mm}  payoff 
 of Alice, the second number denotes the one of Bob and third number denotes one of Charlie.
\includegraphics[scale=0.8,clip]{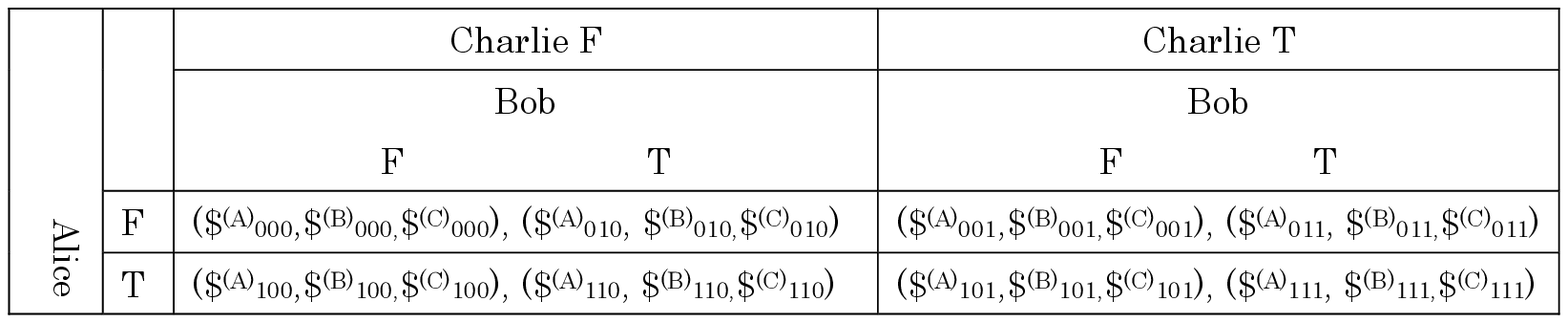} \\
%%Fig.6. Total population $M$ for $n=200$  (left) and $n=1000$  generation (right) in $10$-$th$. 
\end{center}　

%%%%%%%%%%%%%%%%%%%%%
\renewcommand{\thefootnote}{\fnsymbol{footnote}}
%%%%%%%%%%%%%%

%%%%%%%%%%%%%%%%%%%%%%%%%%%%%%%%%%%%%%%%
\section{Protocol for Key Distribution}
%%%%%%%%%%%%%%%%%%%%%%%%%%%%%%%%%%%%%%%%
\hspace{5mm} 
The proposed protocol is based on the three player quantum game. 
Basic idea is that three players can know the full information of game, other players' payoffs and strategies applied,  
when Alice open some classical information to Bob and Charlie.    
However, she can not  get any useful information from it, even if Eva eavesdrops on the classical information. 

%\section{ Scenario of QKD}
\subsection{Basic Protocol of QKD}
\hspace{5mm} 
First of all, we plainly describe the basic protocol  proposed in this paper and the further details will be given the subsequent subsection. 

Notice that all players see the expression Eq.(8) for the expected payoff because they can evaluate it from the setting of the quantum game 
and it is assumed that Alice, Bob and Charlie see the classical payoff matrix described in Table 1( or Alice can set up the values in Table 1 to her convenience ). 
The basic protocol is given by the following 7 steps.   

\begin{enumerate}
\item Alice  prepares an initial quantum state represented by Eq.(1). 

\item Alice sends the second qubit and the third qubit to Bob and Charlie, respectively,  but keeps the first qubit for herself. 

\item 
After  Bob and Charlie accept their qubits, they and Alice locally manipulate their individual qubits by the unitary operator Eq.(2) and  Eq.(3), respectively. 
Three players can choose a favorite parameter set (strategy) of the unitary operators.

\item After that, Bob and Charlie return their qubits manipulated by their unitary operator to Alice. 

\item Alice performs von Neumann measurement of the payoffs based on the measurement basis Eq.(6) and find the strategies applied by  Bob and Charlie from the results 
 \footnote[1]{It is a hard physical problem that what kind of physical experiment should be concretely conducted in order to determine a computational state shrunken at present.
It is known that it is difficult that we even discriminate a special one from four 2qubit Bell states used well in quantum information science\cite{Grus}.%
        Thus it is beyond the limitation of this paper that author presents the concrete experiment that identifies GHZ with 3qubit.}.  
         
\item Alice conveys some information relative to the game  to Bob and Charlie. 

\item By the information, Bob and Charlie can find all information, including opponents' strategy and payoffs, 
of the game. 

\end{enumerate}  

What information Alice should send is determined on a case-by-case as will be described later on in the concrete. 
By the information, Bob and Charlie can find all information, including opponents' strategy and payoffs, of the game. 
Anyway, everyone has common information. 
If they come to an agreement about the correspondence between parameters applied in this protocol and some digital information each other in advance, 
they can have some common digital information. 
This is the essential scenario for the key sharing conveyed by the common information.  
The outline of this protocol is given in Fig.1.   

We show that the scenario  is actually available in the following subsections.  
What information should Alice open to the public in the scenario?  
How can Bob and Charlie find others' information about the game? 
We will show them by giving concrete expressions for them. 
We investigate things dividing into three cases; non-entangled cases, maximally entangled cases and partially entangled cases.  
Explicit expressions of various quantities are, however,  too complicate to analysis them. 
So we impose  some symmetries or conditions on each case in order to simplify things.  
\begin{center}%%%%%%%%%%%%%%%%%%%%%%%%%%%%%
\includegraphics[scale=0.7,clip]{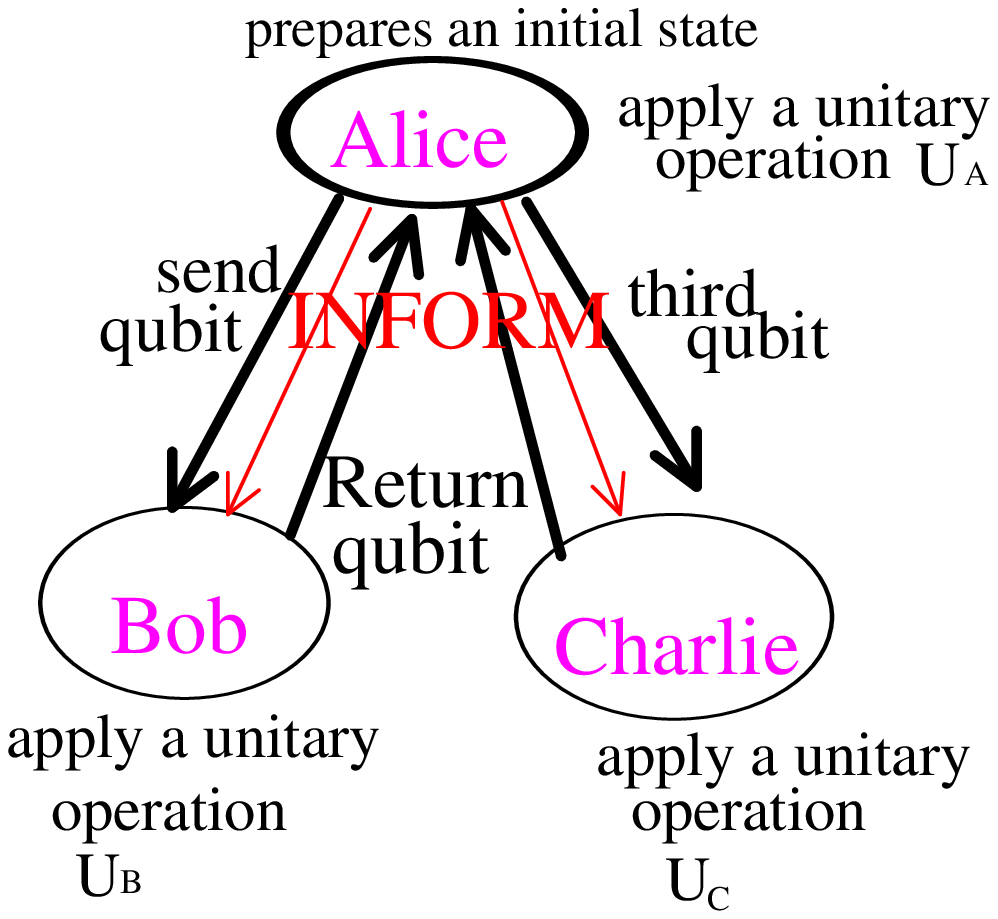} \\
Fig.1  The outline of the protocol　
\end{center}　%%%%%%%%%%%%%%%%%%%%%%%%%%%%%

%%%%%%%%%%%%%%%%%%%%%%%%%%%%%%%%%%%%%%
%\section{ Concrete Scenario of QKD}
%%%%%%%%%%%%%%%%%%%%%%%%%%%%%%%%%%%
\subsection{Non Entanglement Cases} 
\hspace{5mm} 
In this  subsection we consider the case without any entangled states, both the initial state and the computational base.  
We take $\gamma=\delta=0$ which leads to $\eta_1=1$ and  $\eta_2=\xi=0$, and assume that the phase parameters  $\alpha_i=\beta_i=0$. 
  
Then we get the following formula  for the expectation value of the payoff;
\begin{eqnarray}
P^{k} (\theta_k, 0, 0,0,0 )&=&  C_AC_BC_C \$^{(k)}_{000} +  S_AS_BS_C  \$^{(k)}_{111} + C_AC_BS_C\$^{(k)}_{001} + S_AS_BC_C  \$^{(k)}_{110}  \nonumber\\
&&+ S_AC_BC_C  \$^{(k)}_{100} + C_AS_BS_C \$^{(k)}_{011} +  S_AC_BS_C  \$^{(k)}_{101} + C_AS_BC_C \$^{(k)}_{010} \\
&=&C_BC_C D_A^{(k)} + S_BC_C E_A^{(k)} +C_BS_C F_A^{(k)} +S_BS_C G_A^{(k)} \;\mbox{ for  Alice}\\
&=&C_AC_C D_B^{(k)} + S_AC_C E_B^{(k)} +C_AS_C F_B^{(k)} +S_AS_C G_B^{(k)} \;\mbox{ for  Bob}\\
&=&C_AC_B D_C^{(k)} + S_AC_B E_C^{(k)} +C_AS_B F_C^{(k)} +S_AS_B G_C^{(k)} \;\mbox{ for  Charlie},
\end{eqnarray}
where
\begin{eqnarray}
D_A^{(k)} =  C_A\$^{(k)}_{000} +  S_A  \$^{(k)}_{100}&&  
E_A^{(k)} =  C_A \$^{(k)}_{010} + S_A  \$^{(k)}_{110}  \nonumber\\
F_A^{(k)} =  C_A  \$^{(k)}_{001} + S_A \$^{(k)}_{101}    &&
G_A^{(k)} =   C_A  \$^{(k)}_{011} + S_A \$^{(k)}_{111} \\
D_B^{(k)} = C_B\$^{(k)}_{000} +  S_B  \$^{(k)}_{010}  &&
E_B^{(k)} =  C_B \$^{(k)}_{100} + S_B  \$^{(k)}_{110}  \nonumber\\
F_B^{(k)} =  C_B  \$^{(k)}_{001} + S_B \$^{(k)}_{011}   &&
G_B^{(k)} =   C_B  \$^{(k)}_{101} + S_B \$^{(k)}_{111} \\
D_C^{(k)} =  C_C\$^{(k)}_{000} +  S_C  \$^{(k)}_{001}  &&
E_C^{(k)} =   C_C  \$^{(k)}_{100} + S_C \$^{(k)}_{101} \nonumber\\
F_C^{(k)} =  C_C  \$^{(k)}_{010} + S_C \$^{(k)}_{011}    &&
G_C^{(k)} =   C_C \$^{(k)}_{110} + S_C  \$^{(k)}_{111}.  
\end{eqnarray}
Eq.(14)-(17) are the same equations, but they are available expressions for each player, respectively.  
$X^{(k)}_{k\prime}$, where $X=D,E,F,G$ and $k^\prime=$ A,B and C, have 
proper information of player $k^{\prime}$.  

In this stage, we can describe the details of the step 5 in the previous subsection.
\begin{enumerate} [{5}-1] 
\item Alice performs a von Neumann measurement of the payoffs for three players by using the measurement basis Eq.(6),
which is a sort of the physical observable.  
Then the final state shrinks to one of measurement bases with the possibility following quantum theory. 
For example, if the state shrinks to $\psi_{000}$,  Alice finds measurement payoffs $P_A =C_A C_B C_C \$^{(A)}_{000}$ %$
as Alice's payoff.  The measured payoff is represented as $P_A$ with subscript $A$ to mark off from the expected payoff $P^A$. 

\item Alice measure Bob's payoff and Charlie's one in the same manner. They can be simultaneously measured, because their operators are commutable each other. 
Then Alice estimates the ratio $P_A:P_B:P_C$. This value teaches Alice the shrunken state ($\psi_{000}$ in the present example). 
Notice this is possible only when Alice or a planner of the initial payoff matrix needs to alter the ration of $\$^{(k)}_{abc}$ at every set of $\{ abc\}$. %$
Thus Alice can see to which state the final state shrinks from the ratio, 
if the ratio of the payoffs $ \$^{(A)}_{abc}:\$^{(B)}_{abc}:\$^{(C)}_{abc} $ is different each other for  distinct $a$, $b$, $c$. %%%$ 

\item Alice can find strategies adopted by Bob and Charlie (of course her strategy $C_A$), as Alice knows to which state the measured state shrunk. 
For the example of 5.1, Alice see $C_BC_C$\footnote[2]{If Alice wants to know $\theta_B$ and $\theta_C$, independently, 
all players  have only to play  once again with the same strategies as ones in the first round.  
The same value will be unluckily observed with the probability of 1/8. 
Then one more round is needed but repeating  rounds drastically (exponentially)  reduces such unlucky observation.  }. 
By using these values,  Alice calculate the expected payoffs for three players and informs some information relative to game.        
As mentioned later, there is a case that all goes well only by getting the product value $C_BC_C$.   
 \end{enumerate}

%In general it is needed two rounds of 1$\sim$ 5 in subsection 3.1 for  Alice to find Bob's and Charlies's strategy parameters $\theta_B$ and $\theta_C$,  completely. 
%Then Bob and Charlie must adopt the same strategies as ones in the first round.   
%The same value will be unluckily observed with the probability of 1/8. 
%Then one more round is needed but repeating  rounds drastically  reduces such unlucky observation. 
%By using these values,  Alice calculate expected payoffs for three players and informs some information relative to game.        

It is assumed that all players see the classical payoff values described in Table 1 or Alice informs 
the classical payoff values set up in Table I by her to Bob and Charlie.   
What even more information should Alice convey for Bob and Charlie to get information about opponents' payoffs and strategies? 
Notice that three players see the expression Eq. (8) of the expected payoffs. 
For Bob, unknown data are $P^A$,  $P^B$, $P^C$, $C_A$ and $C_C$ which are strategies of Alice and Charlie (notice that  $\alpha_k=\beta_k=0$ 
and of course Bob knows his strategy $C_B$).  
Taking account of $k=A,B,C$, Bob has three payoff equations given by Eq.(16) for $k=A,B,C$.   
So Alice needs to open two data among them to the public. 
Thus Bob can evaluate other unknown data in principle.   
It, however, is too intricate to get explicit expressions. 
Imposing  some conditions or symmetries will make the expressions simpler but non-trivial. 

Due to the aim, Eq. (15)-(17) is  rewritten as followings;
\begin{eqnarray}
P^{k} (\theta_k, 0,0,0,0 )&=&  C_BC_C (D_A^{(k)}  - E_A^{(k)} - F_A^{(k)} +G_A^{(k)}) +C_B( F_A^{(k)}-G_A^{(k)})
\nonumber \\ &&+C_C(E_A^{(k)} - G_A^{(k)}) + G_A^{(k)}  \;\mbox{ for  Alice} \\
&=&C_AC_C (D_B^{(k)}  - E_B^{(k)} - F_B^{(k)} +G_B^{(k)}) +C_A( F_B^{(k)}-G_B^{(k)})
\nonumber \\ &&+C_C(E_B^{(k)} - G_B^{(k)}) + G_B^{(k)} \;\mbox{ for  Bob}\\
&=&C_AC_B (D_C^{(k)}  - E_C^{(k)} - F_C^{(k)} +G_C^{(k)}) +C_A( F_C^{(k)}-G_C^{(k)})
\nonumber \\ &&+C_B(E_C^{(k)} - G_C^{(k)}) + G_C^{(k)} \;\mbox{ for  Charlie}
\end{eqnarray}
It is natural to classify into the following three cases to simplify the things from Eq.(21)-(23); \\

\begin{eqnarray}
\mbox{Case I:}&& (F_A^{(k)}-G_A^{(k)}=)F_B^{(k)}-G_B^{(k)}=F_C^{(k)}-G_C^{(k)}=0 \nonumber \\ 
&& \$^{(k)}_{001} =\$^{(k)}_{101}, \;\;\;\$^{(k)}_{011} =\$^{(k)}_{111} \mbox{ and  } \$^{(k)}_{010} =\$^{(k)}_{110}, \nonumber \\ 
&&(  \$^{(k)}_{001} =\$^{(k)}_{101}=\$^{(k)}_{011} =\$^{(k)}_{111} \mbox{ and  } \$^{(k)}_{010} =\$^{(k)}_{110}),\\    
\mbox{Case II: }&& (E_A^{(k)} - G_A^{(k)}=)E_B^{(k)} - G_B^{(k)}=E_C^{(k)} - G_C^{(k)}=0 \nonumber \\ 
 &&\$^{(k)}_{100} =\$^{(k)}_{101}=\$^{(k)}_{110} =\$^{(k)}_{111} ,\nonumber \\ 
&&(\$^{(k)}_{100} =\$^{(k)}_{101}=\$^{(k)}_{110} =\$^{(k)}_{111} \mbox{ and  } \$^{(k)}_{010} =\$^{(k)}_{011}),\\ 
\mbox{Case III: }&& \mbox{ Case (I) } \bigwedge \mbox{ Case (II) } \nonumber \\ 
&&\$^{(k)}_{001} =\$^{(k)}_{101} =\$^{(k)}_{010} =\$^{(k)}_{100} =\$^{(k)}_{110} =\$^{(k)}_{111}=\$^{(k)}_{011},   \nonumber \\ 
&&\mbox{(The same relations hold )}
\end{eqnarray}  
where the equations within the parentheses denote the cases  that the conditions are also imposed  on Alice. 
Alice is a special person in the sense that  she  puts together everyone's states, observe the payoffs by making a measurement  
and so can see all $P^k$ and $P_k$, and announce them. 
Thus Alice has fully information about this quantum game in the end.  
In above three cases, we can not take $k=A,B,C$ but should take two among $A,B,C$ as $k$,  because 
 we need to satisfy the condition that the ratio of the payoffs $ \$^{(A)}_{abc}:\$^{(B)}_{abc}:\$^{(C)}_{abc} $ is 
different each other for  distinct $a$, $b$, $c$ when $k=A,B,C$. as mentioned in the step 5.2. %%%$ 
%%%%%%%%%%%%%%%%%%%%%%%%%%%%%%%%%%%%%%%%%%%%%%%%%%%%%%%%%%%%%%%%%%%%%%%%%%%%%%%%%%%%
%%%%%%%%%%%%%%%%%%%%%%%%%%%%%%%%%%%%%%%%%%%%%%%%%%%%%%%%%%%%%%%%%%%%%%%%%%%%%%%%%%%%%%
 %\begin{enumerate} [{5}-1] 
 %\item Alice evaluates her payoff from eq. (16) and Bob's and Charlie's strategies given from the measurement.
  %\end{enumerate} 
%%%%%%%%%%%%%%%%%%%%%%%%%%%%%%%%%%%%%%%%%%%%%%%%%%%%%%%%%%%%%%%%%%%%%%%%%%%%%%%%%%%%%

 In the case I without $F_A^{(k)}-G_A^{(k)}=0$,  
we obtain 
 \begin{eqnarray}
C_A&=&\frac{ (P^B- G_B^{(B)})( E_B^{(A)}- G_B^{(A)} ) -(P^A- G_B^{(A)})( E_B^{(B)}- G_B^{(B)} ) }{(P^A- G_B^{(A)})( D_B^{(B)}- E_B^{(B)} ) -(P^B- G_B^{(B)})( D_B^{(A)}- E_B^{(A)} )  }, \nonumber \\
C_C&=&\frac{P^A- G_B^{(A)}}{ C_A  ( D_B^{(A)}- E_B^{(A)} ) +  E_B^{(A)}- G_B^{(A)}   },  \hspace{4cm} \mbox{ for Bob}\\
C_A&=&\frac{ (P^B- G_C^{(B)})( E_C^{(A)}- G_C^{(A)} ) -(P^A- G_C^{(A)})( E_C^{(B)}- G_C^{(B)} ) }{(P^A- G_C^{(A)})( D_C^{(B)}- E_C^{(B)} ) -(P^B- G_C^{(B)})( D_C^{(A)}- E_C^{(A)} )  }, \nonumber \\
C_B&=&\frac{P^A- G_C^{(A)}}{ C_A  ( D_C^{(A)}- E_C^{(A)} ) +  E_C^{(A)}- G_C^{(A)}   },  \hspace{4cm} \mbox{ for Charlie}.
\end{eqnarray}
Notice that  $ E^{(k)}_A=\$^{(k)}_{111} $,  $ F^{(k)}_A=G^{(k)}_A=\$^{(k)}_{001} $ in this case. %$
These quantities are trivial in the sense that they do not depend on Alice's strategy 
and only depend on the payoff matrix  originally opened to the public. 
 $ X_{\bar{k\prime}}^{k}$ where $\bar{k^\prime}=B,C$, however,  is nontrivial and depends on both payoff matrix and  their respective strategy. 
Thus this condition does not make above expressions trivial. 

When $k$ takes A and B in the Case I, from Eq.(27) and (28), we see that when Alice opens $P^A$ and $P^B$ to the public, Bob and Charlie can find  $C_A$. 
As result, they can get the information  about opponent's strategy, $C_C$ for Bob and $C_B$ for Charlie.   
So Bob and Charlie can find the strategies of all players and evaluate $P^k$. 
Three players come to acquire full information of the quantum game, $P^{k}$ and $C_k$,  since Alice originally observes all states and payoffs.  

When Alice informs  Bob of  $P^A$ and his expected payoff $P^B$, and Charlie of  $P^A$ and his expected payoff $P^C$, 
they can also get full information according to the equations derived from Eq.(22) and (23) for Charlie.  
This case,  however, can not be impossible, because Alice needs $k$ to be A, B and C in the condition of the Case I, which was not allowed. 
%\begin{eqnarray}
%C_A&=&\frac{ (P^C- G_C^{(C)})( E_C^{(A)}- G_C^{(A)} ) -(P^A- G_C^{(A)})( E_C^{(C)}- G_C^{(C)} ) }{(P^A- G_C^{(A)})( D_C^{(C)}- E_C^{(C)} ) -(P^C- G_C^{(C)})( D_C^{(A)}- E_C^{(A)} )  }, \nonumber \\
%C_B&=&\frac{P^A- G_C^{(A)}}{ C_A  ( D_C^{(A)}- E_C^{(A)} ) +  E_C^{(A)}- G_C^{(A)}   },  \hspace{4cm} \mbox{ for Charlie}.
%\end{eqnarray}

There is another case to be considered. 
Alice opens all of her information, $P_A$ and $C_A$. 
Then Bob and Charlie can easily evaluate the strategies of their opponent from Eq.(27) and (28). 
So they can also know full information of this game by similar logic to the former case.  
Everything also does not change in the case I including $F_A^{(k)}-G_A^{(k)}=0$. 

 The case II without $E_A^{(k)}-G_A^{(k)}=0$, we obtain 

 \begin{eqnarray}
C_C&=&\frac{ (P^B- G_B^{(B)})( F_B^{(A)}- G_B^{(A)} ) -(P^A- G_B^{(A)})( F_B^{(B)}- G_B^{(B)} ) }{(P^A- G_B^{(A)})( D_B^{(B)}- F_B^{(B)} ) -(P^B- G_B^{(B)})( D_B^{(A)}- F_B^{(A)} )  }, \nonumber \\
C_A&=&\frac{P^A- G_B^{(A)}}{ C_C  ( D_B^{(A)}- F_B^{(A)} ) +  F_B^{(A)}- G_B^{(A)}   },  \hspace{4cm} \mbox{ for Bob}\\
C_B&=&\frac{ (P^B- G_C^{(B)})( F_C^{(A)}- G_C^{(A)} ) -(P^A- G_C^{(A)})( F_C^{(B)}- G_C^{(B)} ) }{(P^A- G_C^{(A)})( D_C^{(B)}- F_C^{(B)} ) -(P^B- G_C^{(B)})( D_C^{(A)}- F_C^{(A)} )  }, \nonumber \\
C_A&=&\frac{P^A- G_C^{(A)}}{ C_B  ( D_C^{(A)}-F_C^{(A)} ) +  F_C^{(A)}- G_C^{(A)}   },  \hspace{4cm} \mbox{ for Charlie}.
\end{eqnarray}
Notice that  $ E_{\bar{k\prime}}^{(k)}=G_{\bar{k\prime}}^{(k)}= \$^{(k)}_{100} $ in this case. %$
When including $E_A^{(k)}-G_A^{(k)}=0$, one more equation $F_{C}^{(k)}=\$_{010}^{(k)}$ is added to the relations. %$
These quantities are trivial in the sense that they do not depend on three players'  strategies 
and only depend on the payoff matrix  originally opened to the public.   
Only $F_{B}^{(k)}$ and  $ D^{(k)}_{k\prime}$, however,  are nontrivial and depend on both payoff matrix and  their respective strategy. 
Thus this does not make above expressions trivial. 

Then we see that when Alice open $P^A$ and $P^B$ to the public, Bob and Charlie can find  $C_A$. 
As result, they can get the information about the opponent's strategy, $C_C$ for Bob and $C_B$ for Charlie.  
So Bob and Charlie can find the strategies of all players and evaluate $P_C$.  
Three players come to  acquire full information of the quantum game, $P^k$ and $C_k$.   

It is also impossible that Alice informs  Bob of  $P^A$  and his payoff $P^B$, and Charlie of  $P^A$ and his payoff $P^C$, because of the same reason as in the Case I, 
%They can also get full information according to the following equations for Charlie; 
%\begin{eqnarray}
%C_A&=&\frac{ (P^C- G_C^{(C)})( F_C^{(A)}- G_C^{(A)} ) -(P^A- G_C^{(A)})( F_C^{(C)}- G_C^{(C)} ) }{(P^A- G_C^{(A)})( D_C^{(C)}- F_C^{(C)} ) -(P^C- G_C^{(C)})( D_C^{(A)}- F_C^{(A)} )  }, \nonumber \\
%C_B&=&\frac{P^A- G_C^{(A)}}{ C_A  ( D_C^{(A)}- F_C^{(A)} ) +  F_C^{(A)}- G_C^{(A)}   },  \hspace{4cm} \mbox{ for Charlie}.
%\end{eqnarray}

There is another case to be considered. 
Alice opens all of her information, $P_A$ and $C_A$. 
Then Bob and Charlie can easily evaluate the strategy of their opponent from the following Eq.(31) and (32). 

\begin{eqnarray}
C_C&=&\frac{ P^A-  G_B^{(A)}  -C_A  ( F_C^{(A)}- G_C^{(A)} ) }{( C_A( D_B^{(A)}- F_B^{(A)} )  }, \hspace{4cm} \mbox{ for Bob} \\
C_B&=&\frac{P^A- G_C^{(A)}  -C_A  ( F_C^{(A)}- G_C^{(A)} ) }{ C_A  ( D_C^{(A)}- F_C^{(A)} )} ,  \hspace{4cm} \mbox{ for Charlie}.
\end{eqnarray}

They can also know full information of this game by similar logic to the former case. 

 Everything becomes simpler in the case III. 
 From Eq. (21)-(23), we obtain
 \begin{eqnarray}
P^k&=&C_AC_C( D_B^{(k)}- F_B^{(k)} ) +  G_B^{(k)} , \nonumber \\
C_C&=&\frac{P^A- G_B^{(A)}}{ C_A  ( D_B^{(A)}- F_B^{(A)} )   },  \hspace{4cm} \mbox{ for Bob}. \\
P^k&=&C_BC_C( D_C^{(k)}- F_C^{(k)} ) +  G_C^{(k)} , \nonumber \\
C_B&=&\frac{P^A- G_C^{(A)}}{ C_A  ( D_C^{(A)}- F_C^{(A)} )   },  \hspace{4cm} \mbox{ for Charlie}. \\
\end{eqnarray}
Then $E^{(k)}_{k\prime}=G^{(k)}_{k\prime}=F^{(k)}_{k\prime}=\$_{100}$.
 %%%%%%%%%%%%%%%%%%%%%%%%%%%%%%%%%%%%%%%%%%%%%%%$
When Alice opens  $P^A$ and $C^A$ to the public, all players know full information of this game for the same logic as  before.

%%%%%%%%%%%%%%%%%%%%%%%%%%%%%\includegraphics[scale=0.8,clip]{hepfig1new.eps} \\
%%%%%%%%%%%%%%%%%%%%%%%%%%%%%Fig．1. A schematic illustration of $m_j$.　　
%\end{center}　

If the further condition $D_{\bar{k\prime}}^{(k)}=F_{\bar{k\prime}}^{(k)}$ and $E_{\bar{k\prime}}^{(k)}=G_{\bar{k\prime}}^{(k)}$ adding to Case III is imposed,  we have only a trivial result. 
Then we notice that   $\$^{(k)}_{000} =\$^{(k)}_{001}$ and $ \$^{(k)}_{010} =\$^{(k)}_{110}$, $\$^{(k)}_{100} =\$^{(k)}_{101}$ and 
$\$^{(k)}_{111}=\$^{(k)}_{110}$ from $D_B^{(k)}=F_B^{(k)}$ and $E_B^{(k)}=G_B^{(k)}$. %$
Moreover we notice that   $\$^{(k)}_{000} =\$^{(k)}_{011}$ and $ \$^{(k)}_{011} =\$^{(k)}_{001}$, $\$^{(k)}_{111} =\$^{(k)}_{101}$ and 
$\$^{(k)}_{100}=\$^{(k)}_{110}$ from $D_C^{(k)}=F_C^{(k)}$ and $E_C^{(k)}=G_C^{(k)}$. %$
So $\$_{0ab}^{(k)}$ take all the same value and   $\$^{(k)}_{1ab}$ take so. %$
After all, $D_{\bar{k\prime}}^{(k)}=F_{\bar{k\prime}}^{(k)}=\$_{000}^{(k)}$, $E_{C}^{(k)}=G_{C}^{(k)}=\$_{100}^{(k)}$ and $E_{B}^{(k)}=G_{B}^{(k)}=\$^{(k)}_{110}$. 
 %$
Thus all $X_{k\prime}^{(k)}$s' are trivial and have no private information for Bob and Charlie.    

%\begin{center}%%%%%%%%%%%%%%%%%%%%%%%%%%%%%
%%%%%%%%%%%%%%%%%%%%%%%%%%%%%\includegraphics[scale=0.8,clip]{hepfig2new.eps} \\
%%%%%%%%%%%%%%%%%%%%%%%%%%%%%Fig．2. A schematic illustration  of Pool and Kochen Model.　　
%\end{center}　

%\begin{center}%%%%%%%%%%%%%%%%%%%%%%%%%%%%%
%%%%%%%%%%%%%%%%%%%%%%%%%%%%%\includegraphics[scale=0.6,clip]{hepfig3new.eps} \hspace{15mm} \includegraphics[scale=0.6,clip]{hepfig4new.eps} \\
%%%%%%%%%%%%%%%%%%%%%%%%%%%%%Fig.3 $a$-$n$ plot for  $N=10^9$.　　 \hspace{15mm} Fig.4 $d$-$n_i $ plot for  $a=10^{-7}$,  $n=10$ and  \hspace*{50mm} $a=10^{-3}$,  $n=1346$.
%\end{center}

%\begin{center}%%%%%%%%%%%%%%%%%%%%%%%%%%%%%
%%%%%%%%%%%%%%%%%%%%%%%%%%%%%\includegraphics[scale=0.8,clip]{hepfigure4new.eps} \\
%%%%%%%%%%%%%%%%%%%%%%%%%%%%%Fig.5. \hspace{5mm} fig.5a.  $n_1$-$\log P_2$  plot  \hspace{25mm}  fig.5b $\log P_2$-$n_1$ plot with the interval 
%of \\ \hspace{90mm}$P_2 =[0.003333333,0.003333334]$.　　
%\end{center}　

%%%%%%%%%%%%%%%%%%%%%%%%%%%%%%%%%%%%%%%
\subsection{Maximally Entanglement Cases} 
%%%%%%%%%%%%%%%%%%%%%%%%%%%%%%%%%%%%%%%%\cite{Pool} 
\hspace{5mm} We consider the maximally entangled cases in the initial state and the computational base where $\gamma=\delta=\pi/2$. 
Then we see that $\xi=\eta_1=\eta_2=1/2$ and the last term  in Eq.(8) that the coefficient of the term is 
$\frac{1}{8} \sin[\theta_A,\theta_B,\theta_C ]  $ vanishes. 
When $\theta_B=\theta_C=0$, we obtain the trivial expected payoff; 
\begin{equation}
P^{k}= C_A \$_{000}^{(k)} \pm S_A \$_{001}^{(k)}. 
\end{equation}
So all people including eavesdropper Eva can obtain full information of this game  as soon as Alice opens something of  (classical) information of this game  to the public.    
If Alice privately conveys it to Bob and Charlie, Eva can obtain full information by eavesdrops. 
 Since the information is a classical type (note the information such as $P^k$ and $C_k$ is classical), it is difficult to detect  the eavesdrops.

We take $\beta_k=\alpha_k=0$ for simplicity but  $\theta_A\theta_B\theta_C\neq 0 $. 
Then we obtain the following equations for the expectation value of the expected payoff;
\begin{eqnarray}
P^{k} (\theta_k,0,0,\pi/2,\pi/2 )&=&  C_AC_BC_C \$^{(k)}_{000} +  S_AS_BS_C  \$^{(k)}_{111} + C_AC_BS_C\$^{(k)}_{001} + S_AS_BC_C  \$^{(k)}_{110}  \nonumber\\
&&+ S_AC_BC_C  \$^{(k)}_{011} + C_AS_BS_C \$^{(k)}_{100} +  S_AC_BS_C  \$^{(k)}_{010} + C_AS_BC_C \$^{(k)}_{101}, \\
&=&C_BC_C D_A^{\prime(k)} + S_BS_C E_A^{\prime(k)} +C_BS_C F_A^{\prime(k)} +S_BC_C G_A^{\prime(k)}, \;\mbox{ for  Alice},\\
&=&C_AC_C D_B^{\prime(k)} + S_AS_C E_B^{\prime(k)} +C_AS_C F_B^{\prime(k)} +S_AC_C G_B^{\prime(k)}, \;\mbox{ for  Bob},\\
&=&C_AC_B D_C^{\prime(k)} + S_AS_B E_C^{\prime(k)} +C_AS_B F_C^{\prime(k)} +S_AC_B G_C^{\prime(k)}, \;\mbox{ for  Charlie},
\end{eqnarray}
 where
 \begin{eqnarray}
D_A^{(k)} =  C_A\$^{(k)}_{000} +  S_A  \$^{(k)}_{011},&&  
E_A^{(k)} =  C_A \$^{(k)}_{101} + S_A  \$^{(k)}_{110},  \nonumber\\
F_A^{(k)} =  C_A  \$^{(k)}_{001} + S_A \$^{(k)}_{110},   &&
G_A^{(k)} =   C_A  \$^{(k)}_{100} + S_A \$^{(k)}_{111}, \\
D_B^{(k)} = C_B\$^{(k)}_{000} +  S_B  \$^{(k)}_{101},  &&
E_B^{(k)} =  C_B \$^{(k)}_{011} + S_B  \$^{(k)}_{111},  \nonumber\\
F_B^{(k)} =  C_B  \$^{(k)}_{001} + S_B \$^{(k)}_{011},   &&
G_B^{(k)} =   C_B  \$^{(k)}_{010} + S_B \$^{(k)}_{111}, \\
D_C^{(k)} =  C_C\$^{(k)}_{000} +  S_C  \$^{(k)}_{001},  &&
E_C^{(k)} =   C_C  \$^{(k)}_{011} + S_C \$^{(k)}_{010}, \nonumber\\
F_C^{(k)} =  C_C  \$^{(k)}_{101} + S_C \$^{(k)}_{100},    &&
G_C^{(k)} =   C_C \$^{(k)}_{110} + S_C  \$^{(k)}_{111}.  
\end{eqnarray}
 
 By comparing these equations to (14)-(20) in the non-entangled case, 
 we find that both expressions are transferred from one hand to the other by exchanging $ 100\leftrightarrow 011$ and   $010\leftrightarrow 101$. 
 So there is a sort of symmetry in the both cases;
 
\begin{equation}
100 \longleftrightarrow  011 \;\; \mbox{ and } \;\;010 \longleftrightarrow 101.
\end{equation}
 Thus there is no essential difference between 
 maximally entangled case and non-entangled case.  
 
 We give a little comment on the cases of $\alpha_k \neq 0 \neq \beta_k$. 
 In both non-entangled case and maximally entanglement case, the last term in Eq.(8) vanishes. 
 Then both cases with  $\alpha_k \neq 0 \neq \beta_k$  are  linearly transformed each other in the elements of a payoff matrix such as;    
 \begin{eqnarray}
\left( \begin{array}{c} 
\$_{000}^\prime\\
\$_{111}^\prime
\end{array}\right) 
&=&
\left( \begin{array}{@{\,}cc@{\,}} 
\eta_1+\xi\cos(\alpha_A+\alpha_B+\alpha_C )&  \eta_2-\xi\cos(\alpha_A+\alpha_B+\alpha_C )\\
\eta_2-\xi\cos(\beta_A+\beta_B+\beta_C ) &  \eta_1+\xi\cos(\beta_A+\beta_B+\beta_C )
\end{array}\right) 
\left( \begin{array}{c} 
\$_{000}\\
\$_{111}
\end{array}\right), 
\nonumber\\
\left( \begin{array}{c} 
\$_{001}^\prime\\
\$_{110}^\prime
\end{array}\right) 
&=&
\left( \begin{array}{@{\,}cc@{\,}} 
\eta_1+\xi\cos(\alpha_A+\alpha_B-\beta_C )&  \eta_2-\xi\cos(\alpha_A+\alpha_B-\beta_C )\\
\eta_2-\xi\cos(\beta_A+\beta_B-\alpha_C ) &  \eta_1+\xi\cos(\beta_A+\beta_B-\alpha_C )
\end{array}\right) 
\left( \begin{array}{c} 
\$_{001}\\
\$_{110}
\end{array}\right), 
\nonumber\\
\left( \begin{array}{c} 
\$_{100}^\prime\\
\$_{011}^\prime
\end{array}\right) 
&=&
\left( \begin{array}{@{\,}cc@{\,}} 
\eta_1-\xi\cos(-\beta_A+\alpha_B+\alpha_C) &  \eta_2+\xi\cos(-\beta_A+\alpha_B+\alpha_C )\\
\eta_2+\xi\cos(\alpha_A-\beta_B+\beta_C ) &  \eta_1-\xi\cos(\alpha_A-\beta_B+\beta_C )
\end{array}\right) 
\left( \begin{array}{c} 
\$_{100}\\
\$_{011}
\end{array}\right), 
\nonumber\\
\left( \begin{array}{c} 
\$_{101}^\prime\\
\$_{101}^\prime
\end{array}\right) 
&=&
\left( \begin{array}{@{\,}cc@{\,}} 
\eta_1-\xi\cos(\beta_A-\alpha_B+\beta_C )&  \eta_2+\xi\cos(\beta_A-\alpha_B+\beta_C )\\
\eta_2+\xi\cos(\alpha_A-\beta_B+\alpha_C ) &  \eta_1-\xi\cos(\alpha_A-\beta_B+\alpha_C )
\end{array}\right) 
\left( \begin{array}{c} 
\$_{101}\\
\$_{101}
\end{array}\right). 
\end{eqnarray}
 So in both non-entangled case and maximally entangled case, $\alpha_k \neq 0 \neq \beta_k$ does not 
have any crucial  influence on previous results in this paper.

%%%%%%%%%%%%%%%%%%%%%%%%%%%%%%%%%%%%%%%
\subsection{Partially Entangled Cases} 
%%%%%%%%%%%%%%%%%%%%%%%%%%%%%%%%%%%%%%%%\cite{Pool} 
\hspace{5mm} 
We take $\alpha_k=0=\beta_C$, $\beta_A-\beta_B=\pi$  and  $\beta_A+\beta_B=2\pi$ for simplicity but  $\theta_A\theta_B\theta_C\neq 0$. 
There are many equivalent choices of these parameters and the this choice is only one example among them.   
The following discussions show that QKD is feasible even in this special case including further condition added later. 

Then we obtain the following equations for the last term in Eq.(8);
\begin{eqnarray}
P^k (\theta_k, 0,\beta_k,\delta,\gamma )_{last}&=&\frac{1}{8} \sin[\theta_A,\theta_B,\theta_C ] \Bigl\{ 
\sin\bigl( \delta-\gamma \bigr) \sum_{a,b,c \in \{0,1 \} } \$^{(k)}_{abc}(-1)^{(a+b+c)}  \Bigr\}. 
\end{eqnarray}

So we consider the case of $\delta=0$ and $\gamma=\pi/2$, or $\delta=\pi/2$ and $\gamma=0$ as a partially entangled case.  
Under this choice, we obtain 
\begin{eqnarray}
\eta_1=\eta_2&=&\frac{1}{2} , \;\;\;\;\;\; \;\;\;\;\;\; \;\;\;\;\;\; \xi =0, \\ 
P^k (\theta_k, 0,\beta_k,\delta,\gamma )_{last}&=&\pm \sqrt{C_AC_BC_CS_AS_BS_C}  \sum_{a,b,c \in \{0,1 \} } 
\$_{abc}^{(k)}(-1)^{(a+b+c)}, 
\end{eqnarray}
where $\pm$ corresponds to two choices of $\delta$ and $\gamma$. 
From Eq.(48), Eq. (8) is rewritten as follows;  
\begin{eqnarray}
P^k (\theta_k,0,\beta_k,\delta,\gamma )&=&  \frac{1}{2} \Big\{ 
\$^{(k)}_{000} \Big(  c_Ac_Bc_C \pm s_As_Bs_C \Big)^2 +\$^{(k)}_{111} \Big(  c_Ac_Bc_C \mp s_As_Bs_C \Big)^2
\nonumber \\
&+& \$^{(k)}_{001} \Big(  c_Ac_Bs_C \mp s_As_Bc_C \Big)^2 +\$^{(k)}_{110} \Big(  c_Ac_Bs_C \pm s_As_Bc_C \Big)^2
 \nonumber\\
&+&\$^{(k)}_{100} \Big(  s_Ac_Bc_C \mp c_As_Bs_C \Big)^2 +\$^{(k)}_{011} \Big( s_Ac_Bc_C \pm c_As_Bs_C \Big)^2
 \nonumber\\
&+& \$^{(k)}_{101} \Big(  s_Ac_Bs_C \pm c_As_Bc_C \Big)^2 +\$^{(k)}_{010} \Big( s_Ac_Bs_C \mp c_As_Bc_C \Big)^2 \Big\}, 
\end{eqnarray}
where 
\begin{eqnarray}
c_k= \cos(\frac{\theta_k}{2}),\;\;\;\;\;\; \;\;\;\;\;\; and \;\;\;\;\;\; \;\;\;\;\;\; s_k= \sin(\frac{\theta_k}{2}).
\end{eqnarray}

When Alice chooses her strategy parameter $\theta_ A=\pi/4$,  the above formula is transformed to

\begin{eqnarray}
P^k (\theta_k, 0,\beta_k,\delta,\gamma  )&=&  \frac{1}{4} \Big\{ 
\$^{(k)}_{000}  \cos^2(\frac{\theta_B \mp \theta_C}{2})   +\$^{(k)}_{111} \cos^2(\frac{\theta_B \pm \theta_C}{2})  
\nonumber \\
&+& \$^{(k)}_{001} \sin^2(\frac{\theta_B \mp \theta_C}{2})   +\$^{(k)}_{110} \sin^2(\frac{\theta_B \pm \theta_C}{2})  
 \nonumber\\
&+&\$^{(k)}_{100}\cos^2(\frac{\theta_B \pm \theta_C}{2})   +\$^{(k)}_{011} \cos^2(\frac{\theta_B \mp \theta_C}{2})  
 \nonumber\\
&+& \$^{(k)}_{101} \sin^2(\frac{\theta_B \pm \theta_C}{2})   +\$^{(k)}_{010} \sin^2(\frac{\theta_B \mp \theta_C}{2}). 
\end{eqnarray}

As pointed out in the subsection 3.2, Alice can find $\theta_B$ and $\theta_C$ by carrying out 1$\sim$5 twice in the subsection 3.1.   
Thus Alice can evaluate the expected payoffs for all players.

Furthermore we impose a condition on the payoff matrix to make the analysis simpler. 
\begin{eqnarray}
\$^{(k)}_{111}= \$^{(k)}_{000},\;\;\;\;\;\; \$^{(k)}_{001}= \$^{(k)}_{110},\;\;\;\;\;\; \$^{(k)}_{100}= \$^{(k)}_{011},\;\;\;\;\;\; \$^{(k)}_{101}= \$^{(k)}_{010}, 
\end{eqnarray}
where $k$=$A $ and $C$ due to  the same reason as in the subsection 3.2.  
This is a sort of duality, since this means $ \$_{abc}=\$_{\bar{a}\bar{b}\bar{c}}  $ where $\bar{0}=1$ and  $\bar{1}=0$. %$
We call this duality NOT-duality that also means $F \Longleftrightarrow T$ symmetry.   
Under this NOT-duality, we obtain 
\begin{eqnarray}
P^k (\theta_k, 0,\beta_k,\delta,\gamma )&=&  \Big\{ 
\$^{(k)}_{000} \Big(  C_AC_BC_C + S_AS_BS_C \Big) +\$^{(k)}_{001} \Big(  C_AC_BS_C + S_AS_BC_C \Big) 
  \nonumber\\
&&+ \$^{(k)}_{100} \Big(  S_AC_BC_C + C_AS_BS_C \Big) + \$^{(k)}_{101} \Big(  S_AC_BS_C + C_AS_BC_C \Big)  \Big\} \\
& =&C_BC_C(  \$^{(k)}_{000}+\$^{(k)}_{100} - \$^{(k)}_{001}-\$^{(k)}_{101}) +C_B(  \bar{F}_A^{(k)} -
 \bar{G}_A^{(k)}) \nonumber\\
&&+ C_C(  \bar{E}_A^{(k)} - \bar{G}_A^{(k)} ) +  \bar{G}_A^{(k)} \;\mbox{ for  Alice}, \\
&=&C_AC_C ( \$^{(k)}_{000}+\$^{(k)}_{101} - \$^{(k)}_{001}-\$^{(k)}_{100}) +C_C(  \bar{E}_B^{(k)}- \bar{G}_B^{(k)})  \nonumber\\
&&+ C_C(  \bar{F}_B^{(k)} - \bar{G}_B^{(k)} ) + \bar{G}_B^{(k)}  \;\mbox{ for  Bob},\\
&=& C_AC_B ( \$^{(k)}_{000}+\$^{(k)}_{001} - \$^{(k)}_{101}-\$^{(k)}_{100}) +C_B(  \bar{F}_C^{(k)}- \bar{G}_C^{(k)})  \nonumber\\
&&+ C_A(  \bar{E}_C^{(k)} - \bar{G}_C^{(k)} ) + \bar{G}_C^{(k)}\;\mbox{ for  Charlie}.
\end{eqnarray}
Here Bob and Charlie do not  necessarily know $C_A$ and  we introduced the following symbol like the previous cases;
 \begin{eqnarray}
\bar{D}_A^{(k)} =  C_A\$^{(k)}_{000} +  S_A  \$^{(k)}_{100}&&  
\bar{E}_A^{(k)} =  C_A \$^{(k)}_{101} + S_A  \$^{(k)}_{001},  \nonumber\\
\bar{F}_A^{(k)} =  C_A  \$^{(k)}_{001} + S_A \$^{(k)}_{101}    &&
\bar{G}_A^{(k)} =   C_A  \$^{(k)}_{100} + S_A \$^{(k)}_{000}, \\
\bar{D}_B^{(k)} = C_B\$^{(k)}_{000} +  S_B  \$^{(k)}_{101}  &&
\bar{E}_B^{(k)} =  C_B \$^{(k)}_{100} + S_B  \$^{(k)}_{001},  \nonumber\\
\bar{F}_B^{(k)} =  C_B  \$^{(k)}_{001} + S_B \$^{(k)}_{100}   &&
\bar{G}_B^{(k)} =  C_B  \$^{(k)}_{101} + S_B \$^{(k)}_{000}, \\
\bar{D}_C^{(k)} =  C_C\$^{(k)}_{000} +  S_C  \$^{(k)}_{001}  &&
\bar{E}_C^{(k)} =  C_C  \$^{(k)}_{101} + S_C \$^{(k)}_{100}, \nonumber\\
\bar{F}_C^{(k)} =  C_C  \$^{(k)}_{100} + S_C \$^{(k)}_{101}    &&
\bar{G}_C^{(k)} =   C_C \$^{(k)}_{001} + S_C  \$^{(k)}_{100}.  
\end{eqnarray}
The expressions of these equations are changed each others under the following transformations;
\begin{eqnarray}
\mbox{Bob} \Longleftrightarrow \mbox{Charlie} &:& \;101 \longleftrightarrow 001
\;\;\;\;\;\;X_B \longleftrightarrow X_C,   \\
\mbox{Bob} \Longleftrightarrow \mbox{Alice} &:& \;100 \longleftrightarrow 101\;\;\;\;\;\;X_B \longleftrightarrow X_A,\\
\mbox{Alice} \Longleftrightarrow \mbox{Charlie} &:& \;100 \longleftrightarrow 001\;\;\;\;\;\;X_A \longleftrightarrow X_C.  
\end{eqnarray}

For example, when Alice opens $C_A=\cos^2(\pi/8)$ $(\theta_A=\pi/4)$ and $P_A$ to the public, Bob and Charlie can find  the strategies of their opponents, respectively;
 \begin{eqnarray}
C_C&=&\frac{P^A- C_A  ( F_B^{(A)}- G_B^{(A)} )+  G_B^{(A)}}{ C_A (\$_{000} + \$_{101} - \$_{100}-\$_{001})+ ( E_B^{(A)}- G_B^{(A)} )   },  \hspace{1cm} \mbox{ for Bob}. \\
C_B&=&\frac{P^A- C_A  ( F_C^{(A)}- G_C^{(A)} )+  G_C^{(A)}}{ C_A (\$_{000} + \$_{001} - \$_{100}-\$_{101})+ ( E_B^{(A)}- G_B^{(A)} )   },   \hspace{1cm} \mbox{ for Charlie}. 
\end{eqnarray}
As result, they find full information of this game. 

Even if Alice opens the expectation values of $P^A$ and $P^B$ based on her observation, Bob and Charlie can also find full information of this game. 
However, the expressions are too complicate to describe them and such too complicated results are not so available for realistic QKD. 
We consider the following symmetric cases for $k=A$ and $B$;
 \begin{eqnarray}
\mbox{(I) }& \bar{E}_B^{(k)}=\bar{G}_B^{(k)}&, \$^{(k)}_{100}= \$^{(k)}_{101} \mbox{ and }  \$^{(k)}_{001}=\$^{(k)}_{000},    \\
\mbox{(II) }& \bar{F}_B^{(k)}=\bar{G}_B^{(k)}&, \$^{(k)}_{001}= \$^{(k)}_{101} \mbox{ and }  \$^{(k)}_{100}=\$^{(k)}_{000}, \\
\mbox{(III) }& \bar{E}_B^{(k)}=\bar{F}_B^{(k)}=\bar{G}_B^{(k)}&, \$^{(k)}_{001}= \$^{(k)}_{101} = \$^{(k)}_{100}=\$^{(k)}_{000}. 
\end{eqnarray}

Case (I);\\
we obtain  
 \begin{eqnarray}
C_C&=&\frac{P^A- \$_{110}+C_B  ( \$_{100}-\$_{001} )}{  (2C_B-1)  (\$_{100}-\$_{001})  },  \hspace{1cm} \mbox{ for Bob}. \\
C_B&=&\frac{P^A- \$_{110}+C_C  ( \$_{100}-\$_{001} )}{  (2C_C-1)  (\$_{100}-\$_{001})  }.   \hspace{1cm} \mbox{ for Charlie}. 
\end{eqnarray}
As result, Bob and Charlie can find full information of this game when only $P^A$ is opened to the public. 

Case (II);\\
we obtain   
 \begin{eqnarray}
C_C&=&\frac{P^A- \$_{000} -C_B  ( \$_{100}-\$_{000} )}{  (2C_B-1)  (\$_{100}-\$_{000})  },  \hspace{1cm} \mbox{ for Bob}. \\
C_B&=&\frac{P^A- \$_{000}+C_C  ( \$_{100}-\$_{001} )}{  (2C_C-1)  (\$_{100}-\$_{001})  }.   \hspace{1cm} \mbox{ for Charlie}. 
\end{eqnarray}
So essentially this case is as same as the case (I). 
Knowing $P^A$ makes all players find full information of this game.  
Thus knowing $P^A$ is only needed to hold full information of the game in common in both Case(I) and (II). 
That such economical point can be  realize is a notable feature in the partially entangled case.

Case (III); 
we only obtain a trivial result;   
\begin{equation}
P^{k}= G^{k}_{k\prime} =1.
\end{equation}

\subsection{More Protocol }
\hspace{5mm} 
We make the QKD protocol described in the subsection 3.1 more definite  so as to be  consummate one. 
The procedure given in the subsection 3.1 has to be repeated many times to distribute a key with a large bit to all players. 
First of all, it is  imperative that three players  ahead recognize the correspondence between  the information of the game and some natural numbers;
\begin{eqnarray}
C_k^{(r)}  &\Longleftrightarrow&  m_k^{(r)}  \;\;\; \mbox{ for  $2r$-$th$  round}, \nonumber \\
P^{k(r)}  &\Longleftrightarrow&  n_k^{(r)}  \;\;\; \mbox{ for  $2r$-$th$  round}, ,
\end{eqnarray}
where $k$ is $A$ or $B$ or $C$, and $m_k^{(r)}$ and $n_k^{(r)}$ are some natural numbers, respectively.  
Thus every player obtains a common number 
\begin{equation}
m_B^{(1)} m_C^{(1)} n_B^{(1)} n_C^{(1)} m_B^{(2)} m_C^{(2)} n_B^{(2)} n_C^{(2)} \cdots m_B^{(r)} m_C^{(r)} n_B^{(r)} n_C^{(r)}
\end{equation}
as a key information by repeating the procedure given in the subsection 3.1  $2r$ times.   
We can obtain a key number with  double the length of  the round number.  
%\begin{center}%%%%%%%%%%%%%%%%%%%%%%%%%%%%%
%\includegraphics[scale=0.8,clip]{hepfig6new.eps} \\
%Fig 7. A schematic diagram of the propagation model. 
%\end{center}　

%\begin{center}%%%%%%%%%%%%%%%%%%%%%%%%%%%%%
%\includegraphics[scale=0.8,clip]{hepfig7new.eps} \\
%Fig 8. $q$-$C_i$ plot for $K=200$ and $\bar{n}=1000$.  
%\end{center}　
\section{Brief Comment of Robustness for  Eavesdrop }
\subsection{Phase Damping Model for  Eavesdropper }
\hspace{5mm} 
There may be  an eavesdropper, Eva, in a quantum line from Alice to Bob or Charlie. 
She may perform a measurement on the qubit that Alice (Bob or Charlie) transmits to Bob or Charlie (Alice). 
We follow Ramzan and Khan \cite{Ramz2} in the discussion  as to security against bugging.

An action of measurement performed by Eva on the qubit can be modeled as the action of phase damping channel \cite{Niel}. 
 After measurement by Eva, the quantum state with 1 qubit that Alice transmitted to Bob and Charlie is transformed into 
\begin{equation}
\rho_1 =\sum_{i=0}^{2} A_i\rho_{in} A^\dagger_i,
\end{equation}
 where $A_0=\sqrt{p}|0><0|$, $A_1=\sqrt{p}|1><1|$ and $A_2=\sqrt{1-p}\hat{I}$ with the identity operator $\hat{I}$ are the Kraus operators\cite{Niel}.  
These operators are the same as ones in the phase flip channel whose channel flips the state of a qubit from $|0>$ to $|1>$ with 
probability $p$\cite{Niel}.     

This can be extended to 
 \begin{equation}
\rho_N =\sum_{i=0}^{2} A_{i_1} \otimes A_{i_2} \cdots   \otimes A_{i_N} \otimes \rho_{in} A^\dagger_{i_N} \otimes \cdots A^\dagger_{i_2} \otimes A_{i_1}^\dagger
\end{equation}
 for $N$ qubits, where the Kraus operator is described as $A_{is} $ for each qubit $s$,  when each qubit is measured. 
 Then the payoff is given by the following replacement in Eq.(8);
\begin{eqnarray}
\xi  &\Longrightarrow&  \xi \mu_p, \nonumber\\
(-1)^{(a+b+c)}   &\Longrightarrow&  \mu_p(-1)^{(a+b+c)},
\end{eqnarray}
where $\mu_p=1-p$. 

 \subsection{Non-Entangled and Maximally Entangled Cases }

\hspace{5mm} In this subsection we focus our attention to the non-entangled case   ($\gamma=\delta=\xi=0$) and maximally entangled case ($\gamma=\delta=\pi/2$ and $\xi=1/2$).  
Then the payoff is obtained  for both cases by 
\begin{eqnarray}
P^{k} (\theta_k, \alpha_k,\beta_k,\gamma,\delta  )&=&  C_AC_BC_C \Big( \eta_1 \$^{(k)}_{000} + \eta_2 \$^{(k)}_{111} +\xi \mu_p (\$^{(k)}_{000}-\$^{(k)}_{111} ) \cos2(\alpha_A +\alpha_B+\alpha_C ) \Big) \nonumber \\
&+& S_AS_BS_C \Big( \eta_2 \$^{(k)}_{000} + \eta_1 \$^{(k)}_{111} -\xi \mu_p (\$^{(k)}_{000}-\$^{(k)}_{111} ) \cos2(\beta_A +\beta_B+\beta_C ) \Big) \nonumber\\
&+& C_AC_BS_C\Big( \eta_1 \$^{(k)}_{001} + \eta_2 \$^{(k)}_{110} +\xi \mu_p (\$^{(k)}_{001}-\$^{(k)}_{110} ) \cos2(\alpha_A +\alpha_B-\beta_C ) \Big) \nonumber\\
&+& S_AS_BC_C \Big( \eta_2 \$^{(k)}_{001} + \eta_1 \$^{(k)}_{110} -\xi \mu_p  (\$^{(k)}_{001}-\$^{(k)}_{110} ) \cos2(\beta_A +\beta_B-\alpha_C ) \Big) \nonumber\\
&+& S_AC_BC_C \Big( \eta_1 \$^{(k)}_{100} + \eta_2 \$^{(k)}_{011} -\xi \mu_p  (\$^{(k)}_{100}-\$^{(k)}_{011} ) \cos2(-\beta_A +\alpha_B+\alpha_C ) \Big) \nonumber\\
&+& C_AS_BS_C \Big( \eta_2 \$^{(k)}_{100} + \eta_1 \$^{(k)}_{011} +\xi  \mu_p (\$^{(k)}_{100}-\$^{(k)}_{011} ) \cos2(-\alpha_A +\beta_B-\beta_C ) \Big) \nonumber\\
&+& S_AC_BS_C \Big( \eta_1 \$^{(k)}_{101} + \eta_2 \$^{(k)}_{010} -\xi \mu_p  (\$^{(k)}_{101}-\$^{(k)}_{010} ) \cos2(\beta_A -\alpha_B+\beta_C ) \Big) \nonumber\\
&+& C_AS_BC_C\Big( \eta_2 \$^{(k)}_{101} + \eta_1 \$^{(k)}_{010} +\xi \mu_p  (\$^{(k)}_{101}-\$^{(k)}_{010} ) \cos2(\alpha_A -\beta_B+\alpha_C ) \Big). 
\end{eqnarray}
In such as the previous cases, taking $\alpha_k=\beta_k=0$, the expected payoffs are obtained for the non-entangled case  and maximally entangled case as the follows; 
\begin{eqnarray}
P^{k} (\theta_k, 0,0,0,0)&=&  C_AC_BC_C  \$^{(k)}_{000}  
+S_AS_BS_C  \$^{(k)}_{111} + C_AC_BS_C \$^{(k)}_{001}  + S_AS_BC_C \$^{(k)}_{110} + S_AC_BC_C \$^{(k)}_{100} \nonumber\\
&+& C_AS_BS_C \$^{(k)}_{011}  
+ S_AC_BS_C  \$^{(k)}_{101} 
+ C_AS_BC_C\$^{(k)}_{010}, \;\mbox{ for  non-entangled case},\\
P^{k} (\theta_k, 0,0,\pi/2,\pi/2)
&=&\frac{1}{2} \Big[  C_AC_BC_C \Big(  \$^{(k)}_{000} + \$^{(k)}_{111} +\mu_p  (\$^{(k)}_{000}-\$^{(k)}_{111} )  \Big) 
+ S_AS_BS_C \Big( \$^{(k)}_{000} +  \$^{(k)}_{111} -\mu_p (\$^{(k)}_{000}-\$^{(k)}_{111} )  \Big) \nonumber\\
&+& C_AC_BS_C\Big( \$^{(k)}_{001} +  \$^{(k)}_{110} + \mu_p  (\$^{(k)}_{001}-\$^{(k)}_{110} ) \Big) 
+ S_AS_BC_C \Big(  \$^{(k)}_{001} +  \$^{(k)}_{110} -\mu_p  (\$^{(k)}_{001}-\$^{(k)}_{110} ) \Big) \nonumber\\
&+& S_AC_BC_C \Big( \$^{(k)}_{100} +  \$^{(k)}_{011} -\mu_p  (\$^{(k)}_{100}-\$^{(k)}_{011} ) \Big) 
+ C_AS_BS_C \Big( \$^{(k)}_{100} + \$^{(k)}_{011} + \mu_p (\$^{(k)}_{100}-\$^{(k)}_{011} )  \Big) \nonumber\\
&+& S_AC_BS_C \Big(  \$^{(k)}_{101} +  \$^{(k)}_{010} -\mu_p  (\$^{(k)}_{101}-\$^{(k)}_{010} )  \Big) 
+ C_AS_BC_C\Big(  \$^{(k)}_{101} +  \$^{(k)}_{010} +\mu_p  (\$^{(k)}_{101}-\$^{(k)}_{010} )  \Big) \Big], \nonumber\\ && \;\hspace{7cm} \mbox{ for  maximally entangled case}. 
\end{eqnarray}
We observe that $\mu_p$ vanishes from $P^{k} $ in the non-entangled case. 
So we can not detect the influence of wiretapping. 
But $P^{k} $ depends on $\mu_p$   in the  maximally entangled case. 
Thus we can detect a wiretapper by comparing two payoffs (one has original value and another has a deviate value from it).  
Notice that when $\alpha_k \beta_k\neq 0 $,   tuning the values of  $\alpha_k $ and $\beta_k $ obscures  $\mu_p$ dependence on the payoff.  
As result, detecting wiretappers is available only at $\alpha_k=\beta_k=0$.  

Though the maximally entangled case and the non-entangled case provided essentially an equivalent way as QKD
 in the previous section,  the latter is not available in the case with wiretappers.

\subsection{Partially Entangled Cases} 
\hspace{5mm}  In this cases,  (i) $\delta=0$ and $\gamma=\pi/2$ or (ii) $\delta=\pi/2$ and $\gamma=0$. 
Moreover $\theta_A\theta_B\theta_C\neq 0$, $\alpha_k=0=\beta_C$, $\beta_A-\beta_B=\pi$ and   $\beta_A+\beta_B=2\pi$  are chosen for simplicity like in the subsection 3.4. 

Then we obtain the following expressions for the last term in Eq. (8);
\begin{eqnarray}
P^k (\theta_k, 0,\beta_k ,\gamma,\delta )_{last}&=&\frac{1}{8} \mu_p \sin[\theta_A,\theta_B,\theta_C ] \Bigl\{ 
\sin\bigl( \delta-\gamma \bigr) \sum_{a,b,c \in \{0,1 \} } \$^{(k)}_{abc}(-1)^{(a+b+c)}  \Bigr\}, \mbox{ for }  (\delta, \gamma)=(0,\pi/2),\nonumber \\
&=&-\frac{1}{8}  \sin[\theta_A,\theta_B,\theta_C ] \Bigl\{ 
\sin\bigl( \delta-\gamma \bigr) \sum_{a,b,c \in \{0,1 \} } \$^{(k)}_{abc}(-1)^{(a+b+c)}  \Bigr\}, \mbox{ for }  (\delta, \gamma)=(\pi/2,0).
\end{eqnarray}

So  the payoff is obtained  by 
\begin{eqnarray}
P^{k} (\theta_k, 0, \beta_k,\gamma,\delta  )&=&\frac{1}{2} \Big[  C_AC_BC_C \Big(  \$^{(k)}_{000} + \$^{(k)}_{111}   \Big) 
+ S_AS_BS_C \Big(  \$^{(k)}_{000} +\$^{(k)}_{111}  \Big) 
+ C_AC_BS_C\Big(  \$^{(k)}_{001} +  \$^{(k)}_{110}  \Big) \nonumber\\
&+& S_AS_BC_C \Big(  \$^{(k)}_{001} +  \$^{(k)}_{110}  \Big)
+ S_AC_BC_C \Big(  \$^{(k)}_{100} +  \$^{(k)}_{011}  \Big) 
+ C_AS_BS_C \Big(  \$^{(k)}_{100} +  \$^{(k)}_{011}  \Big) \nonumber\\
&+& S_AC_BS_C \Big(  \$^{(k)}_{101} + \$^{(k)}_{010} \Big) 
+ C_AS_BC_C\Big( \$^{(k)}_{101} +  \$^{(k)}_{010}  \Big) \Big] +Eq.(81).
\end{eqnarray}
Thus the payoff  depends on $\mu_p=1-p$ only in the case (i) from Eq. (81) and (82).   
In principle,  we can detect an eavesdropper,  only when the parameters are (i) $\delta=0$ and $\gamma=\pi/2$.  
From 4.2 and 4.3, we find that when $\gamma=\pi/2$ where the initial state is an entangled state, we can detect an eavesdropper.

\section{Summary and Consideration}
\hspace{5mm} 
In this paper we proposed a new  QKD method.  
This method is different from the scheme proposed by \cite{Ramz2}, though it essentially takes our ground on three-player  quantum games 
and GHZ state as an entangled state \cite{Gree} is used. 

Alice , Bob and Charlie join the game. 
Alice  prepares the initial quantum state, and sends the second qubit and third qubit to Bob and Charlie, respectively,  but keeps the first qubit for herself. 
After  Bob and Charlie accept their qubits, they locally manipulate their individual qubits by some unitary operator, respectively. 
Three players can choose a favorite parameter set of the unitary operators. 
After that, Bob and Charlie return their qubit manipulated by their unitary operators to Alice. 
Alice performs a measure of the total state (3 qubits)to determine the payoffs, and conveys some information drived from von Neumann measurement to Bob and Charlie or opens to the public.   
By the information, Bob and Charlie can find  opponent's strategy and payoff of the game. 
Thus everyone has common information. 
There are not any arbiters in our scheme, since existence of an arbiter increases the risk of wiretapping. 
For it is difficult to detect wiretapping,  when an arbiter sends classical information.  

We  investigated by  dividing  our protocol into three cases, non-entangled cases, maximally entangled cases and partially entangled cases, to analyze it.  
 We found that  non-entangled cases and maximally entangled ones are essentially equivalent, since they are  converted by a linear transformation each other. 
On the contrary, the partially entangled case has a little particular property and  produces a sort of dense coding method.   

Lastly we discussed robustness for eavesdrop so that  
we showed that though maximally entangled case and  non-entangled case provided essentially equivalent way as QKD,  the latter is not available in the case there are eavesdroppers. 
The effect of eavesdropping disappears from the payoff in the non-entangled case. 
In partially entangled case, we find that we can detect an eavesdropper by choosing some suitable parameter for $\delta$ and $\gamma$, especially $\gamma = \pi/2$,  in principle.  
So this case gives a  robust protocol.  
As summary, we showed an entangled initial state ($\gamma \neq 0$) gives  robust protocols in the all cases of this paper.

\end{document}